# Optimization of Quadratic Forms: NP Hard Problems : Neural Networks

Garimella Rama Murthy,
Associate Professor,
International Institute of Information Technology,
Gachibowli, HYDERABAD,
AP, INDIA

## ABSTRACT

In this research paper, the problem of optimization of a quadratic form over the convex hull generated by the corners of hypercube is attempted and solved. Some results related to stable states/vectors, anti-stable states/vectors ( over the hypercube ) are discussed. Some results related to the computation of global optimum stable state ( an NP hard problem ) are discussed. It is hoped that the results shed light on resolving the $P \neq NP$ problem.

## 1. Introduction:

Constrained / unconstrained optimization problems arise in various areas of human endeavour. Researchers developed various optimization techniques that are routinely applied to solve problems in science, engineering, economics etc. Broadly optimization techniques encompass linear/non-linear programming, integer programming etc, problems. Hopfield proposed a recurrent neural network which acts as an associative memory [Hop]. He reasoned that the network acts as a local / global optimization device for computing the local/global optima of quadratic energy function ( associated with the neural network dynamics ).

Bruck et.al showed that the problem of finding the global optimum stable state is equivalent to finding the minimum cut in the graph corresponding to the Hopfield neural network [BrB]. Thus, solving the minimum cut problem in an undirected graph ( an NP hard problem ) is equivalent to global optimization of the associated quadratic form. Several efforts are made to solve this problem.

The author in his research efforts formulated and solved the problem of optimizing a quadratic form over the convex hull generated by the corners of unit hypercube. This result and the related ideas are documented in Section 2. In Section 3, the relationship between minimum cut computation, neural networks and NP hard problems is discussed. Using the relationship between eigenvalues / eigenvectors and stable values / stable vectors, several properties of local optimum vectors ( over the unit hypercube ) are discussed in Section 4.

Finally some contributions are made towards solving the NP-hard problem of computing the global optimum stable state of a Hopfield neural network. These are documented in Section 5.

## 2. Optimization of Quadratic Forms Over Hypercube:

In this section, we consider the problem of maximization of quadratic form ( associated with a symmetric matrix ) over the corners of binary, symmetric hypercube. Mathematically, this set is specified precisely as follows:

$$S = \{\bar{X} = (x_1, x_2, \ldots, x_N) : x_i = \pm 1 \text{ for } 1 \leq i \leq N \}\ldots(2.1)$$

From now onwards, we call the above set simply as hypercube.
This optimization problem arises in a rich class of applications. This problem is the analogue of the maximization over the hypersphere of quadratic form associated with a symmetric matrix. Rayleigh provided the solution to the optimization problem on the unit hypersphere.

A necessary condition on the optimum vector lying on the unit hypersphere is now provided. This Theorem is the analogue of the maximization over the hypersphere of a quadratic form associated with a symmetric matrix. The following Theorem and other associated results were first documented in [Rama1].

**Theorem 1:** Let $\bar{B}$ be an arbitrary N x N real matrix. From the standpoint of maximization of the quadratic form i.e. $u^T B u$ on the hypercube, it is no loss of generality to assume that **B** is a symmetric matrix with zero diagonal elements. If $u$ maximizes the quadratic form $u^T B u$, subject to the constraint that $|u_i| \leq 1$ for $1 \leq i \leq N$ ( i.e. $\bar{u}$ lies on the hypercube ), then

$$u = sign\,(C u), \quad \ldots\ldots(2.2)$$

where $C = \frac{1}{2}(B + B^T)$ with all the diagonal elements set to zero.

**Proof:** Any arbitrary matrix $B$ can be split into symmetric and skew-symmetric components i.e.

$$C = \frac{1}{2}(B + B^T) \text{ and } \frac{1}{2}(B - B^T)\ldots\ldots\ldots\ldots(2.3)$$

Since the quadratic form associated with the skew symmetric part (matrix) is zero, as far as the optimization of quadratic form is concerned, there is no loss of generality in restricting consideration to symmetric matrices.

- It is now shown that as far as the current optimization problem is concerned, we can only consider symmetric matrices with zero diagonal elements.

Consider the quadratic form $u^T C u$, where the vector $u$ lies on the boundary of the hypercube. Since $u$ lies on the boundary, the quadratic form can be rewritten in the following form:

$$u^T C u = Trace(C) + \sum_{\substack{i=1 \\ i \neq j}}^{N} \sum_{j=1}^{N} u_i C_{ij} u_j \quad \ldots(2.4)$$

Since the Trace (C) is a constant, as far as the optimization over the hypercube is concerned, there is no loss of generality in restricting consideration to a matrix $\widetilde{C}$ whose diagonal elements are all set to zero.

- In the above discussion, we assumed that the optimum of quadratic form over the convex hull of hypercube occurs on the boundary. It will be reasoned in the following discussion.

Now, we apply the discrete maximum principle [ SaW, pp.132 ] to solve the static optimization problem.

Consider a discrete time system
$$Z(k+1) = u(k) \text{ for } k=0,1, \text{ where } u(0) = u. \ldots(2.5)$$
The criterion function to be minimized is given by
$$J^{(0)} = -\frac{1}{2} Z^T(1) \widetilde{C} \, Z(1) = \theta(Z(1), 1). \quad \ldots(2.6)$$
The Hamiltonian is given by

$$H[Z_k, u_k, \rho_{k+1}, k] = \rho_{k+1}^T u(k). \quad \ldots(2.7)$$

From the Discrete maximum principle [ SaW, pp.132 ], since $|u(0)| \leq 1$, the Hamiltonian is minimized when
$$u(0) = -\text{sign}(\rho_1). \quad \ldots(2.8)$$
From the following canonical equation [SaW, pp.133],
$$\rho_1 = \frac{\delta\theta}{\delta Z(1)} = -\widetilde{C} Z(1). \quad \ldots(2.9).$$
Thus, from (2.5), (2.8) and (2.9), we have that

$$u(0) = u = sign\left(\widetilde{C} Z(1)\right) = sign(\widetilde{C} u(0)) = sign(\widetilde{C} u) \ldots(2.10).$$

Thus, the optimal vector $u$ satisfies the necessary condition (2.1) and it lies on the boundary of the hypercube. Q.E.D

**Corollary:** Let $E$ be an arbitrary N x N real matrix. If $u$ minimizes the quadratic form $u^T E u$, subject to the constraint $|u_i| = 1 \text{ for } 1 \leq i \leq n$, then
$$u = -sign(C u), \quad \ldots(2.11)$$

where **C** is the symmetric matrix with zero diagonal elements obtained from **E**.

**Proof:** It may be noted that the same proof as in the above Theorem with the objective function changed from maximization to minimization of quadratic form may be used                                         Q.E.D.

**Remark 0:** The above theorem shows that optimization of a quadratic form over the convex hull generated by the corners of hypercube is equivalent to optimization just over the corners of hypercube ( i.e. local/global optima occur only at the corners of hypercube ).

**Remark 1:** The proof of the above Theorem could be given using other mathematical tools such as non-linear programming ( quadratic optimization ). Also, discrete dynamic programming based proof can be given.

**Remark 2:** It should be noted that the maximization of a quadratic form over a unit hypercube is equivalent to maximization over any hypercube. Countable union of all hypercubes is a subset of the lattice. Thus the optimum over unit hypercube could provide a good approximation to optimization over the symmetric lattice.

**Remark 3:** Now suppose that the second sum in (2.4) does not vanish. Then, utilizing the fact that $u_i u_j = u_j u_i$, it can be rewritten as
$$\sum_{i=2}^{N} \sum_{j=1}^{i} u_i ( b_{ij} + b_{ji} ) u_j = u^T \widetilde{B} u \quad \text{.............(2.12)}$$

where $\widetilde{B}$ is a lower triangular ( could be upper triangular with appropriate summation ) matrix with zero diagonal elements ( Volterra matrix ). Thus, from the standpoint of the optimization over unit hypercube, it is sufficient to consider **B** to be a lower ( upper ) triangular matrix with zero diagonal elements ( Volterra matrix ). Utilization of such a matrix could be very useful in deriving important inferences.

**Remark 4:** An upper bound on the unconstrained objective function is now given through the finite dimensional version of the Cauchy-Schwarz inequality. Let **B u = v**.
$$u^T B u \leq ||u|| \, ||v|| \quad \text{.....(2.13)}$$
where ||.|| denotes the Euclidean norm of the vector. Also equality holds if and only if
$$u = \theta v = \theta B u. \quad \text{........................(2.14)}$$

Thus, the result is in agreement with the Rayleigh's Theorem on optimization of quadratic form on the unit hypersphere.

- A quick argument to show that the maxima always lies on the boundary in the case of positive definite matrices is as follows:
  Suppose not i.e. the extrema ( maxima ) lies inside the n-dimensional hypercube, say at $\tilde{u}$. The value of the quadratic form is given by $\tilde{u}^T B \tilde{u}$. The Euclidean norm of $\tilde{u}$ is clearly less than one. The vector $\frac{\tilde{u}}{||\tilde{u}||}$ which lies on the unit hypersphere gives a larger value for the quadratic form. Thus the claim is true.

**Remark 5**: It is easy to see that a symmetric matrix with zero diagonal elements cannot be positive definite.
  In the following section, we discuss how the problem of maximization of quadratic form naturally arises in connection with the design of Hopfield neural network.

**Remark 6**: As in the case of linear programming, quadratic optimization (considered in this paper) could be carried out using the **interior point methods** guided by the fact that global optimum over the unit hypersphere occurs at the largest eigenvector of a symmetric matrix W. The author is currently investigating this direction [Rama2].

**Remark 7:** The stochastic versions of the problems ( along the lines of Boltzmann machines ) are also currently being investigated by the Author [Rama2] .

**3. Minimum Cut Computation: Neural Networks: NP Hard Problems:**

Researchers in artificial intelligence became interested in developing mathematical models of networks of neurons which can perform some of the functions of biological neural networks. One of the important functions of biological neural network is "associative memory". Hopfield successfully developed one of the earliest models of associative memory.
  Hopfield neural network is a discrete time, non-linear dynamical system that can be represented by a weighted and undirected graph. A weight ( called synaptic weight ) is attached to each edge of the graph and a threshold value is attached to each node ( called neuron ) of the graph. The order of the network , 'n' is the number of nodes

(neurons) in the corresponding graph. Thus a neural network, N of order 'n', is uniquely defined by ( S, T ) where

- S is an n x n symmetric matrix, with $S_{ij}$ being the synaptic weight attached to the edge ( i, j );

- T is a vector of dimension 'n', where $T_i$ denotes the threshold attached to node 'i'.

All the nodes (neurons) in the network can be in one of the two possible states i.e. either +1 or -1. The state space of the network is thus the 'n'-dimensional hypercube (for an 'n'-th order network).

Let the state of the node 'i' at the discrete time instant 't' be denoted by $V_i(t)$. Thus the state of the neural network at time 't' is denoted by the 'n'-dimensional vector V(t). The state of node at time 't+1' is computed in the following manner:

$$V_i(t+1) = Sign\,(J_i(t)) = \begin{cases} +1 & if\ J_i(t) \geq 0 \\ -1 & otherwise \end{cases} \quad \ldots(3.1)$$

where

$$J_i(t) = \sum_{j=1}^{n} S_{i,j} V_j(t) - T_i. \quad \ldots(3.2)$$

The state of the network at time 't+1' i.e. V(t+1) is computed from the current state i.e. V(t) by performing the evaluation (3.1) at a set K of nodes of the network. The set K can be chosen at random or according to some deterministic rule.

The Hopfield neural network operates in various modes. The modes of operation are determined by the method by which the set K is selected at each time instant.

- If the computation in (3.1) is performed at a single node in any time interval, i.e. |K| =1, then the network is operating in the serial mode.

- If |S| = n, then we say that the network is operating in a fully parallel mode.

- All other cases i.e. 1 < |K| < n, will be called the parallel modes of operation

Certain states in the state space are distinguished. For instance

- A State V(t) is called a Stable State if and only if
$$V(t) = Sign\,(\,S\,V(t) - T\,) \quad \ldots\ldots\ldots\ldots(3.3)$$

  i.e no change occurs in the state of the neural network regardless of the mode of operation.

- An important property of the non-linear dynamical system modeling the Hopfield neural network is that it always converges to a stable state when operating in the serial mode and to a cycle of length atmost 2 when operating in a fully parallel mode.

We now formally state the well established convergence Theorem:

**Theorem 2:** Let N = ( S, T ) be a Hopfield neural network with S being the synaptic weight matrix. Then

(A) If the diagonal elements of S are non-negative and the network N is operating in a serial mode, the network will converge to a stable state (i.e there are no cycles in the state space ).

(B) If N is operating in the fully parallel mode, the network will always converge to a stable state or to cycle of length 2

**Remark 8:** The main idea in the proof of the both the parts of the theorem is to define an 'energy function' ( as in Lyapunov stability theory ) and to show that this energy function is non-decreasing when the state of the network changes. Since the energy function is bounded from above, the energy will converge to some value.
The next step in the proof is to show that constant energy implies convergence to stable state in case (A) and to a cycle of length atmost 2 in case (B). The energy function that can be used in the proof of Theorem 2 is

$$E(t) = V^T(t)\, S\, V(t) - 2\, V^T(t)\, T \quad ...……………..(3.4)$$

Thus, the Hopfield neural network, when operating in a serial mode will always get to a stable state that corresponds to a local maximum of the energy function. This result suggests the use of the neural network as an optimization unit for implementing a local search algorithm for finding a maximal value of the energy function. Clearly every optimization problem whose objective function is similar to the quadratic form type expression in (3.4) can be mapped to a Hopfield neural network that will perform a search for its optimum. Let us consider one such problem that arises in graph theory. The required graph-theoretic background is summarized below

- Any graph is associated with two sets: (i) set of vertices and (ii) set of edges. In the case of a weighted graph, the set of weights can be specified through the associated symmetric matrix of weights. Now we formally define minimum cut in the undirected graph in the following:

**Definition:** Let $G = (V, E)$ denote an undirected graph with the associated symmetric matrix of weights, W. Consider a subset U of V and $\tilde{U}$ denote the complement of the set U i.e. $\tilde{U} = V - U$. The set of edges each of which is incident at a node in U and a node in $\tilde{U}$ is called a cut in G. The weight of a cut is the sum of weights of its edges. A minimum cut of a graph is a cut with minimum weight.

The following Theorem proved in [BrB] summarizes the equivalence between the problem of finding minimum cut in a graph and maximizing the energy function of a Hopfield neural network.

**Theorem 3:** Let $N = (S,T)$ be a Hopfield neural network with all the thresholds being zero i.e $T = 0$. The problem of finding a state V for which the energy E is maximum is equivalent to finding a minimum cut in the graph corresponding to N.

In theoretical computer science, it is well known that the problem of computation of minimum cut in an undirected graph is an NP-hard problem i.e from the computational complexity viewpoint, these problems are believed to be intractable. The equivalence in the above Theorem shows that the problem of determination of global optimum stable state of a Hopfield neural network is also an NP-hard problem. Thus, our central goal in this paper is to explore the problem of efficient computation of global optimum stable state of a Hopfield neural network.

In the following section, we explore the connections between eigenvalues/eigenvectors and stable values/stable vectors.

**4. Stable Values, Stable Vectors / States:**

Just like, eigenvalues and eigenvectors arise naturally when the optimization (maximization) of a quadratic form is carried out over the unit hypersphere, stable values and stable vectors of a matrix naturally arise when the maximization is carried out over the unit hypercube.

As discussed earlier, as far as the optimization of the quadratic form is concerned, there is no loss of generality in restricting consideration to quadratic form associated with the corresponding symmetric matrix ( i.e the corresponding symmetric matrix ).

- Also, as discussed in Theorem 1, there is no loss of generality in assuming all the diagonal elements of the matrix A to be zero when the optimization is carried out over the unit hypercube.

Thus, we have the following definition.

**Definition:** A vector **x** in $R^n$ is called a **stable vector** if it stiasfies
$$\mathbf{x} = \text{sign}(\mathbf{M}\mathbf{x}) \quad (\text{M is symmetric})$$
and the corresponding value of the quadratic form is called the **stable value.** Similarly a vector satisfying
$$\mathbf{x} = -\text{sign}(\mathbf{M}\mathbf{x}) \quad (\text{M is symmetric})$$

is called an **anti-stable vector** and the corresponding value of the quadratic form is called the **anti-stable value** .

**Remark 9**: The corresponding definitions dealing with the Hermitian matrix associated with a complex valued ( entries ) matrix are discussed in [Rama 2]. Also, similar definition naturally arise when the optimization is carried out with respect to the $L^p - norm.$

**Remark 10:** It may be relevant to count only the linearly independent vectors which sastisfy $\mathbf{x} = \text{Sign}(\mathbf{M}\mathbf{x})$ as the stable vectors ( similarly anti-stable vectors ). It should be noted that unlike the eigenvectors, the stable vectors / anti-stable vectors may not exist. In the sequel, we mention few examples.
In the following, various facts related to stable values and stable vectors and their cardinality are summarized.

- Trivially, there are atmost $2^n$ stable states/vectors. For instance, when the connection matrix is the identity matrix, all vectors on the hypercube are stable states and the stable value is 'n'. More generally, in the case of a positive definite diagonal matrix, it is easy to see that all the corners of a hypercube are stable vectors with the stable value equal to the weighted sum of the non-negative eigenvalues, where the weights are +1 or -1 corresponding to the components of the stable vector. The global optimum stable vector in this case is the "all-ones" vector denoted by $\bar{e}$. Also, if atleast one eigenvalue of the diagonal matrix is negative, it is again easy to see that no stable vector exists i.e. there is no solution to $\mathbf{u} = \text{Sign}(\mathbf{M}\mathbf{u})$. Furthermore, all the corners of the hypercube are 3anti-stable vectors in the case of a diagonal matrix whose diagonal entries are all negative.

- It is easy to see that, if **y** is a stable state / vector, **-y** is also a stable state/vector. Since
$$x^T M x = (-x^T) M (-x),$$
it is evident that there are atmost $2^{n-1}$ stable values.

- A corner **x** of the hypercube can be mapped to a point **y** on the hypersphere through the transformation

$$y = \frac{X}{\sqrt{n}} = \left(\frac{X_1}{\sqrt{n}}, \frac{X_2}{\sqrt{n}} \ldots \ldots \frac{X_n}{\sqrt{n}}\right).$$

Thus, if a corner of the hypercube is also an eigenvector corresponding to the positive maximum eigenvalue $\mu_{max}$, then that vector provides the maximum stable value, since the corresponding vector on the hypersphere provides the maximum value over all the vectors on the hypersphere, not just the finitely many projections of the corners.

If a corner of the unit hypercube **f** is also an eigenvector corresponding to a positive eigenvalue i.e.
$$\mathbf{M f} = \mu \mathbf{ f,}$$
Then it is also a stable vector, since
$$\text{Sign}(\mathbf{M f}) = \mathbf{f}.$$
Similarly, if a corner of hypercube is also an eigenvector corresponding to a negative eigenvalue, then it is an anti-stable vector.

Having discussed the facts related to stable/ anti-stable vectors / states, we now discuss the NP-Hard problem of computation of global optimum stable state / vector in the following section.

## 5. Global Optimum Stable State Computation:

In view of the results in Section 3, we first consider interesting special cases where the minimum cut computation in an undirected graph can be solved easily.

- Suppose the synaptic weight matrix ( connection matrix ) is non-negative. It is easy to see that the "all-ones" vector i.e. **e** = [ 1, 1, ....,1 ] is the maximum stable state / vector attaining the maximum possible stable value which is equal to the sum of elements of the connection matrix. In the case of non-negative matrices whose rows sum upto a fixed value $\alpha$, the vector **e** is a maximal eigenvector as well as maximal stable vector / state with the corresponding eigenvalue $\alpha$ ( and the corresponding stable value equal to 'n $\alpha$', where 'n' is the dimension of the connection matrix ). It should be noted that similar result can be given with respect to the non-positive matrices.

- Hopfield considered the storage of "s" orthogonal vectors on the hypercube in the associative memory synthesized using the neural network. Specifically let $\{ J_n \}_{n=1}^{s}$ be the orthogonal vectors on the hypercube. It is easy to reason that the following synaptic weight matrix, W, synthesized using those vectors definitely has the 's' vectors as the stable states
$$W = \sum_{n=1}^{s}(J_n J_n^T - I).$$
It is easy to see that
$$W J_l = (N - s) J_l \; for \; 1 \leq l \leq s.$$

Thus, such a connection matrix of the Hopfield associative memory has multiple eigenvalues at '( N –s )' and '-(N-s)' ( N being the dimension of the connection matrix W ). Thus, all the 's' vectors on the hypercube are the maximum stable states/vectors with the stable value being N ( N - s ). In otherwords, the global optimum stable state can be explicitly specified.

- Since, the connection matrix W ( synaptic weight matrix ) is a symmetric matrix, it is always diagonalizable i.e

$$\boldsymbol{W = P\,D\,P^T = \sum_{j=1}^{N} \mu_j \, \overline{f_j}\, \overline{f_j^T}}, \text{ where}$$

**D** is a diagonal matrix with the eigenvalues on the diagonal. The columns of **P** are the right eigenvectors of the connection matrix **W**. Also the matrix **P** is an orthogonal matrix since the right eigenvectors of **W** are orthonormal. Thus, we have the following interesting Lemma

**Lemma 1 :** If one eigenvector of symmetric synaptic weight matrix **W** lies on the hypercube, all other eigenvectors lie on the hypercube.

**Proof:** Proof follows from the fact that the eigenvectors are orthonormal. By Gram-Schmidt orthogonalization procedure, if one eigenvector lies on the hypercube, all other eigenvectors will also lie on the hypercube. Q.E.D.

Thus, in the case of such a connection matrix, the eigenvector corresponding to the largest eigenvalue, $\mu_{max}$ ( that lies on the hypercube ) will necessarily be the stable state corresponding to the maximum stable value $N\,\mu_{max}$. This result naturally initiates the study of matrices all of whose eigenvectors lie on the hypercube [Rama2].

**Remark 11:**
From the above Lemma and well known facts from linear algebra, if one eigenvector doesnot lie on the hypercube, all other eigenvectors also donot lie on hypercube.

- Eliminating the above cases, we now consider a matrix with positive as well as negative entries ( including zeroes ) all of whose eigenvalues donot lie on the hypercube.
The following Lemma is very helpful in the search for a polynomial time algorithm for computing the minimum cut in a graph or equivalently the maximization of a quadratic form associated with the symmetric matrix over the symmetric binary hypercube.

**Lemma 2:** If $y$ is an arbitrary vector on hypercube that is projected onto the unit hypersphere and $x_0$ is the eigenvector of symmetric matrix W corresponding to the maximum eigenvalue ( on the unit hypersphere ), then we have that

$$y^T M y = \mu_{max} + 2\,\mu_{max}(y - x_0)^T x_0 + (y - x_0)^T M (y - x_0)$$

**Proof:** Let $y$ be a vector on the hypercube that is projected onto the hypersphere. Also, let $x_0$ be the eigenvector of the symmetric synaptic weight matrix associated with the maximum eigenvalue. Hence the quadratic form associated with $y$ can be expressed in the following manner.

$$y^T M y = (y - x_0 + x_0)^T M (y - x_0 + x_0)$$
$$= (y - x_0)^T M (y - x_0) + x_0^T M x_0 + 2(y - x_0)^T M x_0$$

Utilizing the fact that $x_0$ is the eigenvector corresponding to the maximum eigenvalue ( maximal eigenvector ) i.e. $M\,x_0 = \mu_{max}\,x_0$ and that $x_0$ lies on the unit hypersphere, that

$$y^T\,M\,y = \mu_{max} + 2\,\mu_{max}(y - x_0)^T\,x_0 + (y - x_0)^T\,M\,(y - x_0).$$

**Q.E.D.**

**Remark 12**: Since, by Rayleigh's theorem, it is well known that the global optimum value of a quadratic form on the unit hypersphere is the maximum eigenvalue i.e. $\mu_{max}$, it is clear that for all corners of the hypercube projected onto the unit hypersphere, we must necessarily have that

$$2\,\mu_{max}(y - x_0)^T\,x_0 + (y - x_0)^T\,M\,(y - x_0) \leq 0.$$

The goal is to choose a **y**, such that the above quantity is as less negative as possible ( so that the value of quadratic form is as close to $\mu_{max}$ as possible.

- **Heuristic Algorithm for Computation of Global Optimum Stable State of a Hopfield Neural Network:**

Step 1: Suppose the right eigenvector corresponding to the largest eigenvalue of M is real ( i.e. real valued components ). Compute such an eigenvector, **z**

Step 2: Compute the corner, **T** of hypercube from **z** in the following manner:
$$T = \text{Sign}(z).$$

Step 3: Using **T** as in the initial condition ( vector ), run the Hopfield neural network in the serial mode of operation.

In view of Lemma2 and above remark , the claim is that the global optimum stable state is reached through the above procedure.

**Note:** Clearly, the above algorithm fails when the eigenvector corresponding to the largest eigenvalue is complex valued.

We now propose a method which reduces the computational complexity of the method of computing the global optimum stable state.

**Claim:** Given a linear block code, a neural network can be constructed in such a way that every local maximum of the energy function corresponds to a codeword and every codeword corresponds to a local maximum.

**Proof:** Refer the paper by Bruck et.al [BrB].

It has been shown in [BrB] that a graph theoretic code is naturally associated with a Hopfield network ( with the associated quadratic energy function ). The local and global optima of the energy function are the codewords.

**Goal: To compute the global optimum stable state ( i.e. global optimum of the energy function ) using the associated graph theoretic encoder.**
To achieve the goal, once again the largest real eigenvector ( if it is real valued ) is utilized as the basis for determining the information word that will be mapped to a global optimum stable state/ codeword ( using the associated graph theoretic encoder ).

- **Probabilistic Formulation of the Problem of Computation of Global Optimum Stable State:**
  From basic linear algebra, it is well known that a symmetric matrix, W is diagonalizable and can be wriiten in the following form:

$$W = P\, D\, P^T,$$

where P is an orthogonal matrix i.e. $P^T = P^{-1}$ and D is a diagonal matrix containing the eigenvalues of W on the diagonal. Now project all the corners of hypercube onto the unit hypersphere. Thus, if **x** is a corner of the N-dimensional hypercube, the corresponding projected point on the unit hypersphere is given by

$$y = \frac{x}{\sqrt{N}}.$$

Since the columns of the orthogonal matrix **P** are orthonormal, we have that
   **y** = **P C**, where **C** is an Nx1 vector. Since **y** lies on the unit hypersphere, we have that

$$\boldsymbol{C^T C} = 1 = \sum_{i=1}^{N} c_i^2$$

Thus, we have that

$$y^T W y = C^T P^T P D P^T P C = C^T D C = \sum_{i=1}^{N} c_i^2 \mu_i ,$$

where $\mu_i$'s are the eigenvalues of W. Thus, from the above expression, we infer that the value of quadratic form is the expectation of the random variable which assumes the values $\mu_i$'s with the corresponding probabilities $c_i^2$. Each corner of hypercube that is projected onto the unit hypersphere, leads to certain expectation value. The goal is to find the global optimum value of the expectation and the vector at which it is assumed. This constitutes a probabilistic formulation of the problem being considered.

6. **Conclusions:**

   In this research paper, it is shown that optimizing the quadratic form over the convex hull generated by the corners of hypercube is equivalent to optimization over just the corners of hypercube. The relationship between minimum cut computation, neural networks and NP-hard problems is summarized. Several properties of stable states / anti-stable states are summarized. Some results related to the computation of global optimum stable state are discussed.

**References:**

[BrB]  J. Bruck and M. Blaum,"Neural Networks, Error Correcting Codes and Polynomials over the Binary Cube," IEEE Transactions on Information Theory, Vol.35, No.5, September 1989.

[Hop]  J.J. Hopfield, "Neural Networks and Physical Systems with Emergent Collective Computational Abilities," Proceedings of National Academy Of Sciences, USA Vol. 79, pp. 2554-2558, 1982

[Rama1]  G. Rama Murthy,"Optimal Signal Design for Magnetic and Optical Recording Channels, " Bellcore Technical Memorandum, TM-NWT-018026, April 1st , 1991

[Rama2]  G. Rama Murthy,"Efficient Algorithms for Computation of Minimum Cut in an Undirected Graph,"

[SaW]  A.P. Sage and C.C. White,"Optimum Systems Control," Prentice Hall Inc, 1977